\documentclass{aa}
\usepackage{graphicx,natbib,color,bm,url}
\graphicspath{{./fig/}{./png/}}

\newcommand{\EQ}{\begin{equation}}
\newcommand{\EN}{\end{equation}}
\newcommand{\EQA}{\begin{eqnarray}}
\newcommand{\ENA}{\end{eqnarray}}
\newcommand{\eq}[1]{(\ref{#1})}

\newcommand{\Eq}[1]{Eq.~(\ref{#1})}

\newcommand{\Eqss}[2]{Eqs.~(\ref{#1})--(\ref{#2})}

\newcommand{\Sec}[1]{Sect.~\ref{#1}}

\newcommand{\Fig}[1]{Fig.~\ref{#1}}
\newcommand{\FFig}[1]{Figure~\ref{#1}}

\newcommand{\Figs}[2]{Figs.~\ref{#1} and \ref{#2}}

\newcommand{\Tab}[1]{Table~\ref{#1}}

\newcommand{\bra}[1]{\langle #1\rangle}

\newcommand{\meanrho}{\overline{\rho}}

{}
{}
{}

\newcommand{\meanSSSS}{\overline{\mbox{\boldmath ${\mathsf S}$}} {}}

{}
{}
{}
{}
{}
{}
{}
{}
{}
{}
{}
{}
{}
{}
{}
{}
{}
\newcommand{\meanUU}{\overline{\bm{U}}}

{}

\newcommand{\meanA}{\overline{A}}
\newcommand{\meanB}{\overline{B}}

\newcommand{\meanU}{\overline{U}}

\newcommand{\means}{\overline{s}}

\newcommand{\meanP}{\overline{P}}

\newcommand{\meanT}{\overline{T}}

{}

{}
{}
{}

\newcommand{\nnn}{\hat{\bm{n}}}

\newcommand{\meanAA}{{\overline{\bm{A}}}}
\newcommand{\meanBB}{{\overline{\bm{B}}}}
\newcommand{\meanJJ}{{\overline{\bm{J}}}}

\newcommand{\nullvector}{{\bf0}}

\newcommand{\xx}{\bm{x}}

\newcommand{\bb}{\bm{b}}

\newcommand{\grav}{\bm{g}}
\newcommand{\BB}{\bm{B}}

\newcommand{\UU}{\bm{U}}
\newcommand{\FF}{\bm{F}}

\newcommand{\uu}{\bm{u}}

\newcommand{\nab}{{\bm{\nabla}}}

\newcommand{\nabad}{\nabla_{\rm ad}}

\newcommand{\DD}{{\rm D} {}}

\newcommand{\dd}{{\rm d} {}}

\newcommand{\const}{{\rm const}  {}}

\def\Pm{\mbox{\rm Pr}_M}

\def\cp{c_{\rm p}}
\def\cv{c_{\rm v}}

\def\cs{c_{\rm s}}

\def\qpz{q_{\rm p0}}
\def\qp{q_{\rm p}}

\def\betap{\beta_{\rm p}}

\def\betastar{\beta_{\star}}
\def\Peff{{\cal P}_{\rm eff}}

\def\kf{k_{\rm f}}

\def\Hp{H_{\rm p}}

\def\sigmaSB{\sigma_{\rm SB}}

\def\urms{u_{\rm rms}}

\def\nut{\nu_{\rm t}}

\def\nuT{\nu_{\rm T}}

\def\etat{\eta_{\rm t}}

\def\etaT{\eta_{\rm T}}

\def\Beq{B_{\rm eq}}
\def\Beqz{B_{\rm eq0}}
\def\tautd{\tau_{\rm td}}

\def\half{{\textstyle{1\over2}}}

\def\onethird{{\textstyle{1\over3}}}

\newcommand{\G}{\,{\rm G}}

\newcommand{\K}{\,{\rm K}}
\newcommand{\g}{\,{\rm g}}
\newcommand{\s}{\,{\rm s}}

\newcommand{\ks}{\,{\rm ks}}
\newcommand{\cm}{\,{\rm cm}}

\newcommand{\km}{\,{\rm km}}

\newcommand{\kms}{\,{\rm km/s}}

\newcommand{\Mm}{\,{\rm Mm}}

\newcommand{\yapj}[3]{ #1, {ApJ,} {#2}, #3}

\newcommand{\yapjl}[3]{ #1, {ApJ,} {#2}, #3}

\newcommand{\yan}[3]{ #1, {Astron.\ Nachr.,} {#2}, #3}

\newcommand{\yana}[3]{ #1, {A\&A,} {#2}, #3}

\newcommand{\ypf}[3]{ #1, {Phys.\ Fluids,} {#2}, #3}

\newcommand{\yjetp}[3]{ #1, {Sov.\ Phys.\ JETP,} {#2}, #3}

\newcommand{\ymn}[3]{ #1, {MNRAS,} {#2}, #3}

\newcommand{\ysph}[3]{ #1, {Solar Phys.,} {#2}, #3}

\newcommand{\ypre}[3]{ #1, {Phys.\ Rev.\ E,} {#2}, #3}

\newcommand{\yjour}[4]{ #1, {#2}, {#3}, #4}

\newcommand{\ybook}[3]{ #1, {#2} (#3)}

\title{Spontaneous flux concentrations from the negative effective
magnetic pressure instability beneath a radiative stellar surface}
\titlerunning{The negative magnetic pressure instability with radiation}
\author{B. Perri\inst{1} \and A. Brandenburg\inst{2,3,4,5}}
\authorrunning{Perri \& Brandenburg}
\institute{
DSM/IRFU/SAp, CEA-Saclay and UMR AIM, CEA-Universit\'e Paris 7, F-91191 Gif-sur-Yvette, France
\and
JILA and Department of Astrophysical and Planetary Sciences, University of Colorado, Boulder, CO 80303, USA
\and
Department of Astronomy, AlbaNova University Center,
Stockholm University, 10691 Stockholm, Sweden
\and
Nordita, KTH Royal Institute of Technology and Stockholm University, 10691 Stockholm, Sweden
\and
Laboratory for Atmospheric and Space Physics, University of Colorado, Boulder, CO 80303, USA
}

\date{\today,~ $ $Revision: 1.86 $ $}

\begin{document}
\abstract{
The formation of sunspots requires the concentration of magnetic flux
near the surface.
The negative magnetic pressure instability (NEMPI) might be a possible
mechanism for accomplishing this, but it has mainly been studied in simple
systems using an isothermal equation of state without a natural free surface.
}{
We study NEMPI in a stratified Cartesian mean-field model where turbulence
effects are parameterized.
We use an ideal equation of state and include radiation transport,
which establishes selfconsistently a free surface.
}{
We use a Kramers-type opacity with adjustable exponents chosen such
that the deeper layers are approximately isentropic.
No convection is therefore possible in this model, allowing us to
study NEMPI with radiation in isolation.
We restrict ourselves to two-dimensional models.
We use artificially enhanced mean-field coefficients to allow NEMPI
to develop, making it therefore possible to study the reason why it
is much harder to excite in the presence of radiation.
}{
NEMPI yields moderately strong magnetic flux concentrations a certain
distance beneath the surface where the optical depth is unity.
The instability is oscillatory and in the form of upward travelling waves.
This seems to be a new effect that has not been found in earlier models
without radiative transport.
The horizontal wavelength is about ten times smaller than what
has previously been found in more idealized isothermal models.
}{
In our models, NEMPI saturates at field strengths too low to explain
sunspots.
Furthermore, the structures appear too narrow and too far beneath the surface
to cause significant brightness variations at the radiative surface.
We speculate that the failure to reproduce effects resembling sunspots may
be related to the neglect of convection.
}
\keywords{Radiative transfer -- hydrodynamics -- Sun: atmosphere --
Sun: sunspots
}

\maketitle

\section{Introduction}

Sunspots are a highly intermittent manifestation of strong magnetic flux
concentrations at the solar surface.
The underlying magnetic fields are produced by hydromagnetic turbulence
in the convection zone beneath the surface \citep{BMT04,BMBBT11,KMB12,ABMT15}.
Numerous simulations of turbulence and turbulent convection have
displayed such magnetic field production by a dynamo process
\citep{BS05}.
This alone, however, does not explain the occasional concentration
into spots.
On the other hand, more realistic simulations that include the effects
of strong density stratification near the surface, as well as radiation
and ionization, have been able to demonstrate the appearance of magnetic
spots \citep{SN12}.
Furthermore, it is known that strong density stratification can lead to a
large-scale instability of an initially unstructured random magnetic
field \citep{KRR89,KRR90}.
This instability leads to magnetic flux concentrations and even magnetic
spots \citep{BKR13,WLBKR13} through the negative magnetic pressure
instability (NEMPI).
NEMPI has been associated with sunspot formation by \cite{KRR95,KMR96},
following a series of earlier work on its theoretical foundations
\citep{KMR93,KR94,RK07}.

Numerical simulations have also displayed types of magnetic flux
concentrations that are not straightforwardly associated with NEMPI.
This tends to be the case when the magnetic field is produced by a large-scale
dynamo some distance beneath the surface \citep{MBKR14,JBKMR15,JBMKR16}.
Nevertheless, also in those cases strong stratification was shown to be
essential, as has been demonstrated by comparing with weakly stratified
cases.

In the case of NEMPI, the underlying instability can well be modeled
using mean-field magnetohydrodynamics, where the negative effective
magnetic pressure is parameterized in terms of the mean magnetic field;
see \cite{BRK16} for a review.
In some of those cases there is good quantitative agreement between
direct numerical simulations (DNS) and mean field simulations (MFS),
as has been demonstrated in several papers \citep{KBKMR13,Los2}.

The main difference between MFS and DNS is the inclusion of the
parameterization of the small-scale unresolved motions
$\uu=\UU-\meanUU$ and magnetic fields $\bb=\BB-\meanBB$
in the MFS.
Here, the overbar denotes a suitably defined average, which, in
practice, could be a spatial average.
The evolution equations for $\meanUU$ involve correlations
of the form $\overline{u_i u_j}$ and $\overline{b_i b_j}$ 
that need to be expressed in terms of $\meanUU$ and $\meanBB$.
They are similar to the parameterization in terms of the rate-of-strain
tensor of the mean flow involving turbulent viscosity, but there are
also contributions that are quadratic in $\meanBB$.
Similar parameterizations also exist for the Maxwell stresses in the
momentum equation and the electromotive force in the induction equation.

Before we can think of applying NEMPI to real sunspot formation, we
must begin to address the effects of radiation, ionization, and other
potentially important surface effects.
Here, we focus on radiative transfer.
Radiation has two important effects.
On the one hand, it leads to the establishment of a natural surface
from which most of the observed radiation is emitted and above which
the density drops off sharply.
On the other hand, radiation also leads to the equilibration of
temperature differences between neighboring fluid elements.
Earlier investigations have suggested that this may indeed
be the case and that NEMPI may be difficult to excite in the presence
of radiation \citep{Bar13,BB16}.
This is also the reason why we focus here on mean-field simulations,
because they allow us to artificially exaggerate the effects of NEMPI
by choosing unrealistically large mean-field parameters, which allows
us to study the properties of NEMPI in that case and helps us
determining the conditions under which NEMPI may still operate.

Most of the earlier investigations of NEMPI have been carried out in an
isothermally stratified layer using an isothermal equation of state.
This means that no energy equation was solved.
This was also true in simulations with an outer coronal envelope
\citep{WLBKR13}, where the interface was characterized by a layer
above which the driving of turbulence was turned off.
The aim of the present paper is therefore to study NEMPI
in a simple model with radiative heating and cooling included.

\section{The model}

\subsection{Mean-field equations and radiative transfer}

We consider the mean-field equations in Cartesian coordinates,
but restrict ourselves to including
only the effects of turbulent magnetic diffusion, turbulent viscosity,
and the negative effective magnetic pressure effect, which means that
the ordinary magnetic pressure from the mean field, $\meanBB^2/2\mu_0$,
is modified and becomes $(1-\qp)\meanBB^2/2\mu_0$, where
$\qp=\qp(\meanBB)$ depends on the local magnetic field strength.
We write the mean magnetic field as $\meanBB=\BB_0+\nab\times\meanAA$,
where $\BB_0=(0,0,B_0)$ is an imposed vertical field, and $\meanAA$
is the mean magnetic vector potential.
We thus solve the equations for $\meanAA$, the mean velocity $\meanUU$,
the mean specific entropy $\means$, and the mean density $\meanrho$
in the form
\EQ
{\partial\meanAA\over\partial t}=\meanUU\times\meanBB+\etaT\nabla^2\meanAA,
\label{dAdt}
\EN
\EQ
\meanrho{\DD\meanUU\over\DD t}
=-\nab\left(\meanP-{\qp\meanBB^2\over2\mu_0}\right)
+\meanJJ\times\meanBB+\meanrho\grav+\nab\cdot(2\nuT\meanrho\,\meanSSSS),\;
\EN
\EQ
\meanrho\meanT{\DD\means\over\DD t}=-\nab\cdot(\FF_{\rm rad}+\FF_{\rm conv})
+2\nuT\meanrho\,\meanSSSS^2,
\label{DsDt}
\EN
\EQ
{\DD\ln\meanrho\over\DD t}=-\nab\cdot\meanUU,
\EN
where $\etaT=\eta+\etat$ is the total magnetic
diffusivity consisting of a microphysical and a turbulent value,
$\nuT=\nu+\nut$ is the total viscosity consisting again of a
microphysical and a turbulent value, $\overline{\sf S}_{ij}=\half
(\meanU_{i,j}+\meanU_{j,i})-\onethird\delta_{ij}\nab\cdot\meanUU$
is the traceless rate-of-strain tensor,
$\meanJJ=\nab\times\meanBB/\mu_0$ is the Lorentz force from the mean
fields (without the effects of turbulence that are being parameterized
through $\qp$), $\grav=(0,0,-g)$ is the gravitational acceleration,
$\meanP$ is the mean gas pressure, $\meanT$ is the mean temperature,
$\DD/\DD t=\partial/\partial t+\meanUU\cdot\nab$ is the advective derivative,
$\FF_{\rm rad}$ is the radiative flux, and $\FF_{\rm conv}$ is the
convective flux, but it will be neglected in our present exploratory work.

The radiative flux divergence is obtained by solving the
radiative transfer equations for the intensity $I(\xx,t,\nnn)$
in the gray approximation in the form \citep{Nor82}
\EQ
\nnn\cdot\nab I=-\kappa\meanrho\, (I-S)
\label{RT-eq}
\EN
along a set of rays in different directions $\nnn$, where $\kappa$ is
the opacity and $S=(\sigmaSB/\pi)\,\meanT^4$ is the source function
with $\sigmaSB$ being the Stefan--Boltzmann constant.
The radiative flux divergence is found by integrating \Eq{RT-eq}
over all directions, i.e.,
\EQ
\nab\cdot\FF_{\rm rad}=-\kappa\meanrho\oint_{4\pi}(I-S)\,\dd\Omega,
\label{fff}
\EN

We adopt the equation of state for a perfect gas, i.e.,
$\meanP=({\cal R}/\mu)\meanT\,\meanrho$, where ${\cal R}$ is the
universal gas constant and $\mu$ the mean specific weight.
The mean specific entropy is, up to an irrelevant additive constant,
given by $\means/c_p=(\ln\meanP)/\gamma-\ln\meanrho$, where
$\gamma=\cp/\cv$ is the ratio of specific heats at constant pressure
and constant density, respectively, and ${\cal R}/\mu=\cp-\cv$.
In the following, we take $\gamma=5/3$ which is appropriate
for a monatomic gas and in the absence of ionization.
The pressure scale height, $\Hp=-\dd\ln\meanP/\dd z$,
is then given by $\Hp={\cal R}\meanT/\mu g$.
In the isothermal part near the top, pressure and
density scale heights are equal, i.e., $H_\rho=\Hp$,
where $H_\rho=-\dd\ln\meanrho/\dd z$.
However, in the deeper isentropic parts, we have
$H_\rho=\gamma \Hp$.

\subsection{Parameterizations}

Turbulence effects such as NEMPI depend on the relative importance
of the magnetic field to the equipartition field strength with respect
to the turbulent energy, that is, on $\beta\equiv|\meanBB|/\Beq$.
Here, the equipartition field strength $\Beq$ is given by
$\Beq^2(z)=\mu_0\meanrho\urms^2$.
The effective magnetic pressure is characterized by the
functional form of $\qp=\qp(\beta)$, for which we assume \citep{KBKR12}
\EQ
\qp(\beta)={\qpz\over1+\beta^2/\betap^2}
={\betastar^2\over\betap^2+\beta^2},
\quad\mbox{where $\betastar=\betap\qpz^{1/2}$}.
\label{qpFormula}
\EN
In addition, we have to specify $\etaT$ and $\nuT$, which we assume to
be constant and equal to each other, i.e., we assume the turbulent
magnetic Prandtl number $\Pm=\nuT/\etaT$ to be unity \citep{YBR03}.
We define a fiducial model where we take $\qpz=300$ and $\betap=0.05$.
Earlier work of \cite{KBKMR13} showed that the growth rate is mainly
dependent on the parameter $\betastar$, whose value is then 0.87.
For comparison, \cite{KBKR12} and \cite{KBKMR12} used the parameter
combination $\qpz=40$ and $\betap=0.05$, which then yields about
a third for $\betastar=0.32$.
Our value of $\betastar$ is thus much higher than what has been assumed
before, which should help us to study the effects of radiation in the
development of NEMPI.
Following earlier work of \cite{Bar13} and \cite{BB14},
we assume a Kramers-like opacity law for $\kappa$ of the form
\EQ
\kappa=\kappa_0(\meanrho/\rho_0)^a (\meanT/T_0)^b
\EN
with constant coefficients $\kappa_0$, $\rho_0$, and $T_0$,
and given exponents $a$ and $b$.
The resulting radiative conductivity is then given by
\citep{BB14}
\EQ
K=K_0\,{(\meanT/T_0)^{3-b}\over(\meanrho/\rho_0)^{1+a}}
=K_0\,\left[{(\meanT/T_0)^n\over\meanrho/\rho_0}\right]^{1+a},
\EN
where
\EQ
K_0=16\sigmaSB T_0^3/3\kappa_0\rho_0
\label{K0def}
\EN
is a constant and
\EQ
n=(3-b)/(1+a)
\label{polytropic}
\EN
is, for $n>-1$, related to the polytropic index
of the resulting stratification.
The radiative diffusivity is $\chi=K/\meanrho\cp$.
The optical depth is $\tau(z)=\int_z^\infty\kappa\meanrho\,\dd z'$.
The region where $\tau\ll1$ is optically thin, while the region where $\tau\gg1$
is optically thick, which corresponds to the
convection zone in the Sun; $\tau=1$ represents thus the solar surface.

\subsection{Boundary conditions and numerical aspects}

We adopt impenetrable stress-free vertical field boundary conditions
in the $z$ direction, so the velocity obeys
\EQ
\partial\meanU_x/\partial z=\partial\meanU_y/\partial z=\meanU_z=0
\quad\mbox{on $\;z=0$, $L_z$},
\EN
where $L_z$ is the vertical extent of the computational domain
and the bottom boundary is at $z=0$.
For the magnetic field we adopt the vertical field condition,
\EQ
\partial\meanA_x/\partial z=\partial\meanA_y/\partial z=\meanA_z=0
\quad\mbox{on $\;z=0$, $L_z$}.
\EN
We assume zero incoming intensity at the top, and compute the incoming
intensity at the bottom from a quadratic Taylor expansion of the source
function, which implies that the diffusion approximation is obeyed;
see Appendix~A of \cite{HDNB06} for details.
As in \cite{BB14}, we fix the temperature at the bottom,
\EQ
\meanT=T_0\quad\mbox{on $z=0$},
\EN
while the temperature at the top is allowed to evolve freely.
There is no boundary condition on the density, but since no mass
is flowing in or out, the volume-averaged density is
automatically constant; see Appendix~C of \cite{BB14}.

To reduce the computational expense, we solve \Eqss{dAdt}{fff}
in two spatial dimensions.
In an earlier investigation of NEMPI, \cite{Los1} found that this
simplification can lead to about two times smaller growth rates,
but the qualitative dependencies on various input parameters
were still reproduced correctly.
In the present model, we use either
$288$ or $576$ meshpoints in the $z$ direction.
The number of mesh points in the $x$ direction depends on the domain size
and is constrained such that the mesh spacings $\delta x$ and $\delta z$
are equal in the two directions.
We employ the {\sc Pencil Code}\footnote{
\url{https://github.com/pencil-code}}, where all relevant terms
are readily implemented.
The code uses a high-order finite-difference scheme.
The radiation module was implemented by \cite{HDNB06}.

\subsection{Comparison with the optically thick approximation}
\label{OpticallyThick}

It will be instructive to compare with the more familiar case in which
$F_{\rm rad}$ is computed in the optically thick approximation as
$\FF_{\rm rad}=-K\nab\meanT$ in a domain $0\leq z\leq d$, where $d$ is less
than the $L_z$ used in the general case with full radiative transfer.
At $z=d$, we apply a radiative boundary condition
\EQ
\partial\meanT/\partial z=-\sigmaSB\meanT^4\quad\mbox{(on $z=d$)}.
\label{radbc}
\EN
The value of $d$ is computed from \citep[see Sect.~3.12 of][]{BB14} as
\EQ
d=\cp(T_0-T_1)\,\nabla_{\rm ad}/(g\nabla),
\label{ddef}
\EN
where $\nabla_{\rm ad}=1-1/\gamma$ and $\nabla=1+1/n$
with $n$ given by \Eq{polytropic}, and
\EQ
T_1={\cal K}^{1/4}\,T_0\quad\mbox{with}\quad
{\cal K}={gK_0\over\cp\sigmaSB T_{\rm bot}^4}\,{\nabla\over\nabad},
\label{calK}
\EN
where $K_0$ is given by \Eq{K0def}.
The quantities in \Eqss{ddef}{calK}
are fully determined by the parameters of the radiative model.
We emphasize that the temperature at the top is close to $T_1$,
but it allowed to evolve freely subject to \Eq{radbc}.
Computationally, the optically thick approximation is by about a
factor of two cheaper, but it is more restrictive, because the
values of $d$ and $T_1$ are intimately tied to the choice of
$\kappa_0$ and cannot be varied independently.

\subsection{Scale separation ratio}
\label{ScaleSep}

In our mean-field model, turbulence is parameterized in terms of a
magnetic turbulent diffusivity, which is estimated to be
$\etat=\urms/3\kf$ \citep{SBS08}, where $\kf$ is the wavenumber of the
energy-carrying motions.
We compare this with the reference wavenumber $k_1=2\pi/L_z$
based on our domain of height $L_z$.
We refer to $\kf/k_1$ as the scale separation ratio.
Thus, we have \citep{JBLKR14}
\begin{equation}
\kf/k_1=\urms/3\etat k_1.
\end{equation}
This ratio must be large enough for NEMPI to be excited \citep{BKKR12}.
Early DNS of \cite{BKKMR11}, where NEMPI was excited, used $\kf/k_1=15$,
but with $\kf/k_1=30$, NEMPI became much more pronounced \citep{KBKMR12}.

Following the work of \cite{BB14}, we measure length in $\Mm$,
velocity in $\km\s^{-1}$, and density in $\g\cm^{-3}$.
We choose $L_z=5\Mm$.
We adopt a squared domain, $L_x=L_z$, and assume
for the turbulent small-scale velocity $\urms=1\kms$.
Thus, we have $\etat=5\times10^{-3}\Mm\kms$, so we have $\kf/k_1=53$,
which should be large enough for NEMPI to be excited \citep{BKKR12}.
In some models with larger resolution ($576^2$ meshpoints), we used
$\etat=2\times10^{-3}\Mm\kms$, corresponding to $\kf/k_1=133$;
see \Tab{units} for the conversion of several quantities from code
units to cgs units.
Following earlier work \citep{BKKMR11}, we also define the general
turbulent--diffusive time $\tautd=(\etat k_1^2)^{-1}$.

\begin{table}[t!]\caption{
Units used in this paper and conversion into cgs units.
}\vspace{12pt}\centerline{\begin{tabular}{lll}
\hline\hline
quantities&code units&cgs units\\
\hline
length [$z$]       &    $\Mm$    & $10^8\cm$\\
velocity [$u$]     &   $\km\s^{-1}$    & $10^5\cm\s^{-1} $ \\
time $[t]$, $[\lambda]^{-1}$ &   $\ks$    & $10^3\s $ \\
density {[$\rho$]}&    $\g\cm^{-3} $ &$1\g\cm^{-3}$\\
temperature {[$T$]}     & $\K$ & $1\K$\\
time {[$t$]} & $\ks$ & $10^3\s$ \\
gravity {[$g$]}& $\km^2\s^{-2}\Mm^{-1}$& $10^2\cm\s^{-2}$\\
opacity  {[$\kappa$]}& $\Mm^{-1}\cm^3\g^{-1}$ &
$10^{-8}\cm^2\g^{-1}$\\
diffusivity [$\chi$, $\etat$, $\nut$]& $\Mm\km\s^{-1}$ & $10^{13}\cm^2\s^{-1}$\\
\hline
\label{units}\end{tabular}}\end{table}

\section{Results}

\begin{figure}[t!]\begin{center}
\includegraphics[width=\columnwidth]{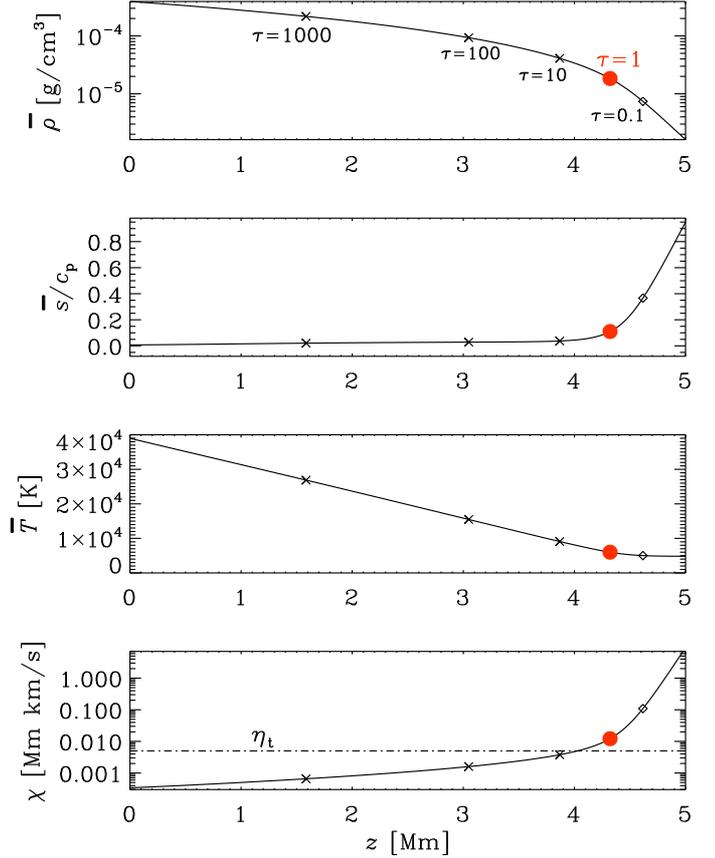}
\end{center}\caption[]{
Stratification of $\meanrho$, $\means/\cp$, $\meanT$, and $\chi$.
The location of $\tau=1$ is marked with a red filled symbol,
while the diamond indicates $\tau=0.1$ and the two crosses
denote $\tau=10$ and $100$.
Here, $\kappa_0=10^7\Mm^{-1}\cm^3\g^{-1}$.
}\label{pstrat_a1b0_288a}\end{figure}

We design the model such that it has an isentropic deeper part.
The stratification in our model is similar to Run~B7 of \cite{BB14}
with $a=1$ and $b=0$, which, as discussed above, yields $n=1.5$.
In particular, we use $\kappa_0=10^7\Mm^{-1}\cm^3\g^{-1}$, which results
in a surface temperature of around $5000\K$.
As in \cite{BB14}, we compute a hydrostatic equilibrium solution
($\uu=\nullvector$) by solving \Eqss{dAdt}{fff} only in the $z$ direction
in one dimension.
The result is shown in \Fig{pstrat_a1b0_288a}, where we
plot the $z$ dependence of $\meanrho$, $\means$, $\meanT$, and $\chi$.
In the deeper parts, where $\tau\gg1$,
$\meanT$ increases linearly with depth and,
because $\means$ is nearly constant in that part,
$\meanrho(z)\propto\meanT^{3/2}$, which is in agreement
with the expected polytropic stratification.
Above the surface, $\meanT(z)$ is approximately constant,
so $\meanrho(z)$ falls off exponentially with height,
as expected for an isothermal stratification.
We begin by discussing in some detail a run with $200\G$, which
will later also be referred to as Run~B$''$; see \Tab{ta:data}.
The presence of an imposed field changes the stratification,
but this change is small: $T$ decreases by $\approx4\K$
for $B_0=200\G$.

\begin{table}[b!]
\caption{Summary of Runs~A--H.
}\begin{center}
\begin{tabular}{llccccccc}
   & $L_x$ & $B_0$ & $\kappa_0$ & $\etat$ & $N$ & $\lambda$ &
$\omega$ & $z_B$ \\
\hline
\hline
A   & $1.25$ & 100 &  $10^7$   & $5\;10^{-3}$ & $288^2$ & 0.011 & 0.79 & 3.0 \\ 
A$'$& $2.5$  & 100 &  $10^7$   & $5\;10^{-3}$ & $288^2$ & 0.009 & 0.89 & 2.5 \\ 
B   & $1.25$ & 200 &  $10^7$   & $5\;10^{-3}$ & $288^2$ & 0.030 & 1.44 & 2.7 \\ 
B$'$& $2.5$  & 200 &  $10^7$   & $5\;10^{-3}$ & $288^2$ & 0.017 & 1.42 & 2.5 \\ 
B$''$&$5.0$  & 200 &  $10^7$   & $5\;10^{-3}$ & $288^2$ & 0.049 & 1.21 & 2.5 \\ 
C$'$& $2.5$  & 500 &  $10^7$   & $5\;10^{-3}$ & $288^2$ & 0.021 & 0.90 & 1.7 \\ 
\hline
D   & $1.25$ & 200 & $2\;10^7$ & $5\;10^{-3}$ & $288^2$ & 0.022 & 1.51 & 2.8 \\ 
E   & $1.25$ & 200 & $5\;10^7$ & $5\;10^{-3}$ & $288^2$ & 0.043 & 1.55 & 3.2 \\ 
\hline
F   & $1.25$ & 200 & $2\;10^7$ & $2\;10^{-3}$ & $576^2$ & 0.094 & 0.70 & 2.7 \\ 
G   & $1.25$ & 200 & $5\;10^7$ & $2\;10^{-3}$ & $576^2$ & 0.101 & 0.83 & 2.8 \\ 
H   & $1.25$ & 200 &  $10^8$   & $2\;10^{-3}$ & $576^2$ & 0.152 & 0.70 & 3.1 \\ 
\end{tabular}
\label{ta:data}
\end{center}
\tablefoot{
All quantities are measured in code units; see \Tab{units}.
In the first group of runs, $B_0$ is varied.
In the second and third groups, $\kappa_0$ is varied,
but in the third one, $\etat$ is also decreased.
}\end{table}

\begin{figure*}[t!]\begin{center}
\includegraphics[width=\textwidth]{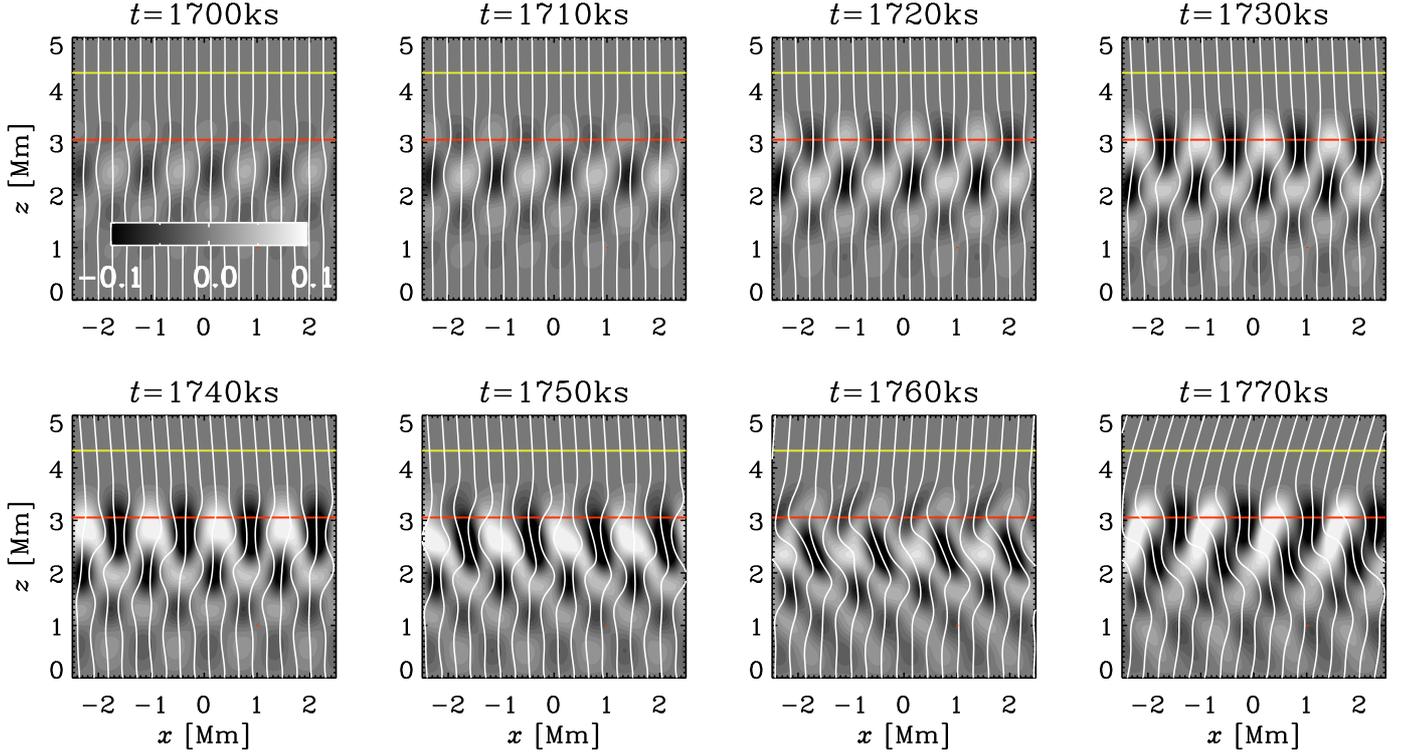}
\end{center}\caption[]{
Gray scale representation of vertical velocity
together with magnetic field lines in white
for a run with $B_0=200\G$ (Run~B$''$).
The yellow and red horizontal lines are the $\tau=1$
and $\tau=100$ surfaces, respectively,
}\label{pslice1700_1770}\end{figure*}

\begin{figure}[t!]\begin{center}
\includegraphics[width=\columnwidth]{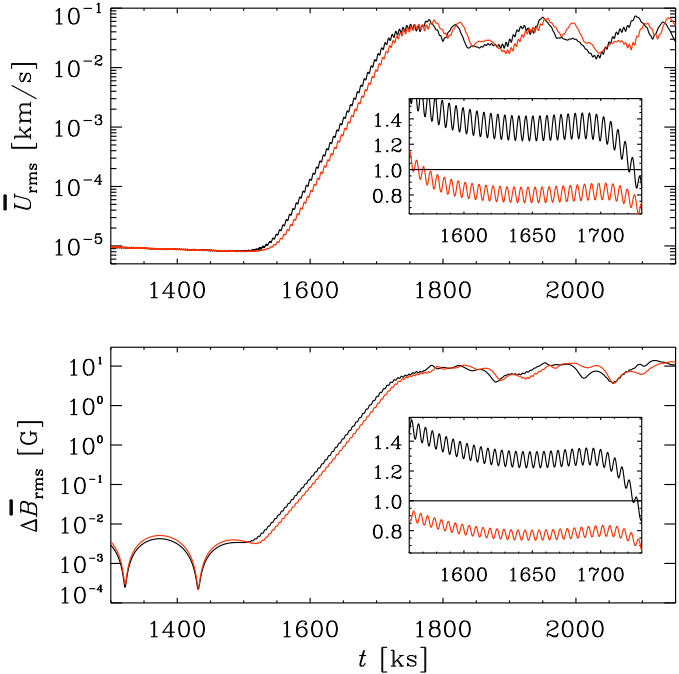}
\end{center}\caption[]{
Evolution of $\meanU_{\rm rms}$ and $\Delta\meanB_{\rm rms}$ for two
runs with different initial seed magnetic field and $B_0=200\G$ in
both cases.
The insets show the compensated functions, $e^{-\lambda t}\meanU_{\rm rms}$
and $e^{-\lambda t}\Delta\meanB_{\rm rms}$, respectively,
where $\lambda$ is the growth rate.
Only the short time interval of exponential growth is shown.
}\label{pts_a1b0_288a}\end{figure}

\subsection{Early evolution into saturation}

In the early phase of the evolution, structures form where
$\max(|\meanBB|)$ is at $z=z_B\approx2.5\Mm$ and a horizontal
wavenumber $k=4\,k_1$; see \Fig{pslice1700_1770}.
These structures gradually move downward, disappear, and new ones form
at $z\approx3\Mm$.
Those structures then also move downward, and so on.
The structures occur well below the $\tau=1$ line and are close to the
$\tau=100$ line.
Here, the photon mean-free path,
\EQ
\ell=(\kappa\rho)^{-1},
\label{elldef}
\EN
is about $0.05\Mm$, while at $\tau=1$, it is about $0.14\Mm$.
The downward motions are associated with a local field enhancements,
as can clearly be seen from field lines getting more concentrated in
some locations.
At later times, the field becomes more irregular, but retains a
typical horizontal wavenumber of $4\,k_1$.
In some cases, however, we found that, in the late nonlinear stage,
$k$ can decrease from four to three.

The growth of structures can be characterized both by the typical
velocities $\meanUU$ in the domain and the departures from the
imposed field $\Delta\meanBB=\meanBB-\BB_0$.
In \Fig{pts_a1b0_288a}, we show for two independent realizations of
Run~B$''$ (with $L_x=L_z$) the evolution of the rms values,
$\meanU_{\rm rms}$ and $\Delta\meanB_{\rm rms}$,
with different seeds for the random initial velocity perturbations.
We clearly see an oscillatory growth of both quantities, as can
also be seen by showing a plot compensated by $\exp(-\lambda t)$,
where $\lambda\approx0.048\ks^{-1}$ is the growth rate, as determined
during the exponential growth phase of the instability.
In the following, we measure the period $P_{\rm osc}$ as the volume-integrated
rms velocity of the mean field and record the frequency $\omega=2\pi/P_{\rm osc}$.
The frequency of the actual (signed) magnetic field is half that value.

The growth rate is independent of the initial seed for the random number
generator, but the detailed nonlinear evolution does depend on it
(compare the lines in each of the panels of \Fig{pts_a1b0_288a}).
This suggests that the evolution of NEMPI is chaotic in the
nonlinear regime.
Animations show that the field lines are constantly swinging back and forth.
This type of time-dependence of NEMPI is new and has not previously been
seen -- neither in isothermal nor in polytropic calculations.
It may therefore be an effect related to the presence of radiation.
The slight apparent difference in oscillation amplitudes of the compensated
plots in the insets is caused by the fact that both have been compensated
by the same factor, but the amplitudes were slightly different
by the time the eigenfunction begins to be established.

\begin{figure}[b!]\begin{center}
\includegraphics[width=\columnwidth]{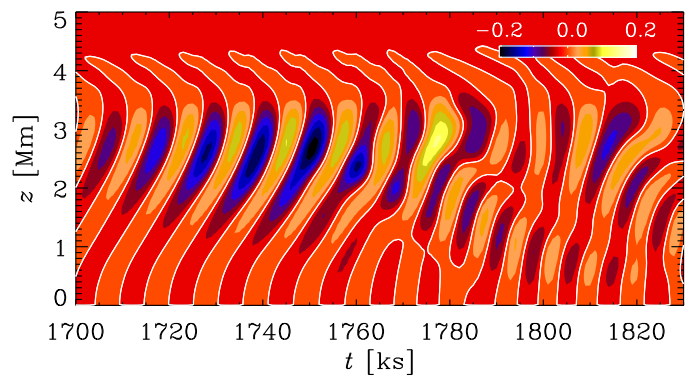}
\end{center}\caption[]{
$\meanU_z$ (color coded) versus $t$ and $z$ for Run~B.
The zero contours are shown in white.
}\label{puz_tz_1700_1900}\end{figure}

The spatio-temporal evolution of NEMPI is seen more clearly in
\Fig{puz_tz_1700_1900}, where we show $\meanU_z(x_\ast,z,t)$ for
$x_\ast=-1.7\Mm$.
This position $x_\ast$ is where $\meanU_z$ has an anti-node during
the linear growth phase.
At $t=1740\ks$, the slope in the $tz$ diagram corresponds to a pattern
speed of about $0.2\km\s^{-1}$, which is small compared with the sound
speed $\cs\approx20\km\s^{-1}$ at $z\approx2.5\Mm$,
and it is also small compared with the turbulent rms velocity of $1\km\s^{-1}$,
but it agrees with the typical NEMPI-produced downflow speeds found earlier for
isothermal NEMPI experiments \citep{BGJKR14}.

\subsection{Dependence on control parameters}

We now consider the dependence of NEMPI on $B_0$, $\kappa_0$, and $\etat$.
We revisit some of these dependencies later in more detail.
Several input and output parameters of our runs are summarized in
\Tab{ta:data}.
Although most of the runs discussed in this paper are performed for
a domain with $L_x=L_z=5\Mm$, several aspects can also be reproduced
in narrower domains with $L_x/L_z=0.5$ and $0.25$.
The growth rate $\lambda$ is rather sensitive to this,
while the oscillation frequency $\omega=2\pi/P_{\rm osc}$ and the position
$z_B$ of the magnetic field maximum are less sensitive.
For Runs~A--C with $\kappa_0=10^7\Mm^{-1}\cm^3\g^{-1}$, the growth rates
are roughly in the range between $\lambda=0.01\ks^{-1}$ and $0.03\ks^{-1}$
and do not seem to be systematically dependent on the value of $B_0$.
This is mainly related to the fact that NEMPI can develop deeper down
as $B_0$ is increased; see \cite{KBKR12}.
This is characterized by the value of $z_B$ given in
\Tab{ta:data}; compare especially with the value for Run~C.
Our results thus confirm that the structures develop at larger
depths when the field becomes stronger.
This is in agreement with earlier work \citep{KBKR12,LBKR14}.

\subsection{Comparison with earlier work}

In units of $\tautd$ (defined in \Sec{ScaleSep}), the growth rate is
$\tilde\lambda\equiv\lambda\tautd=\lambda/\etat k_1^2$,
which is about 6 for Run~B$''$.
However, if we normalize instead by actual horizontal wavenumber
$k$ of the structures, which is 4 times larger than $k_1$ (see
\Fig{pslice1700_1770}), we have $\lambda/\etat k^2\approx0.4$.
This value is rather low and comparable to the value in the first DNS
of \cite{BKKMR11}, where the scale separation ratio was much lower
($\kf/k_1=15$ compared to $53$ in the present case).

\begin{figure}[t!]\begin{center}
\includegraphics[width=\columnwidth]{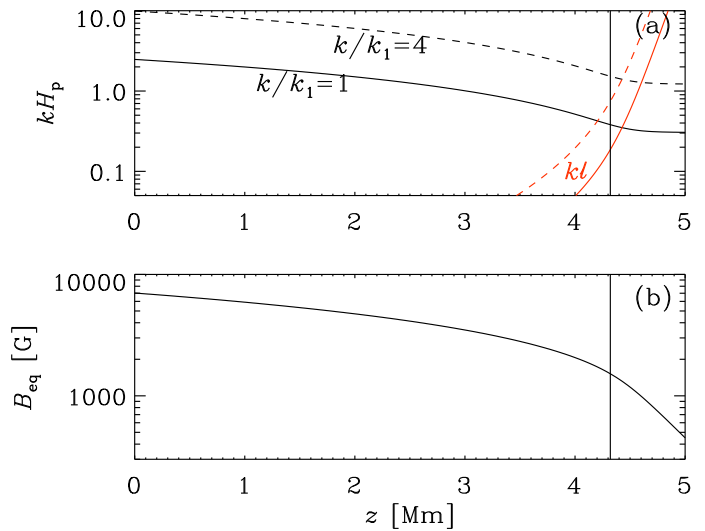}
\end{center}\caption[]{
Vertical dependence of $k\Hp$ for $k/k_1=1$ and 4 (a),
and of $\Beq$ (b).
The vertical lines denote the surface where $\tau=1$.
In panel (a), we also show in red dependence of $k\ell$.
}\label{pHrho_a1b0_288a}\end{figure}

Earlier work using isothermal layers has shown that the horizontal
wavenumber of the instability is comparable to the inverse density
scale height; \cite{KBKMR13} found $kH_\rho=1.1$--$1.5$.
Subsequent work showed that during the nonlinear evolution of the
instability, $kH_\rho$ can decrease from about 0.8 to 0.2.
This has been associated with an inverse cascade-type behavior
\citep{BGJKR14}.
The polytropic simulations of \cite{LBKR14} gave larger values:
$kH_\rho=1$ in the
upper layers and $kH_\rho=2$ in deeper ones; see their Fig.~12.
In the present case, at the height where the instability based on
the absolute field strength is strongest ($z=3\Mm$), and for $k/k_1=4$,
we find $k\Hp=5$; see the dashed line in \Fig{pHrho_a1b0_288a}a.
This is a striking difference between the present models and the earlier ones
using an isothermal equation of state.

\FFig{pHrho_a1b0_288a}b shows that, in the region where NEMPI develops,
the equipartition field strength $\Beq(z)$ is around $3000$--$4000\G$.
This is about 20 times larger than the strength of the imposed field,
which is typical of NEMPI and in agreement with earlier results
\citep{LBKR14,BGJKR14}.

\subsection{Magnetic field dependence}

\begin{figure}[b!]\begin{center}
\includegraphics[width=\columnwidth]{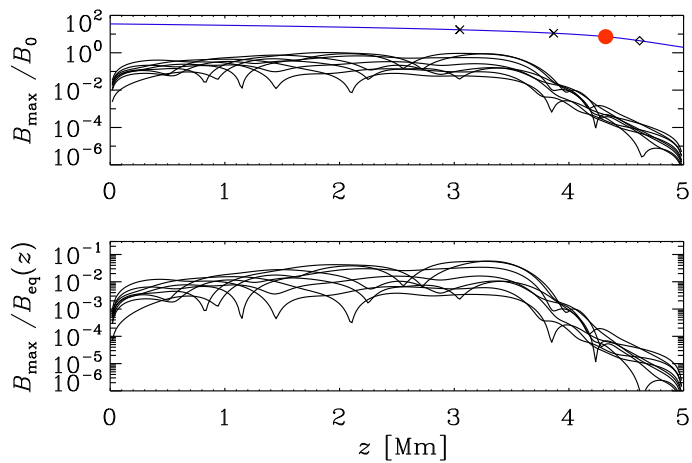}
\end{center}\caption[]{
Vertical dependence of the normalized magnetic field for 
different times in the nonlinear phase for $B_0=200\G$ (Run~B$''$).
The location of $\tau=1$ is marked with a red filled symbol,
while the diamond indicates $\tau=0.1$ and the two crosses
denote $\tau=10$ and $100$.
}\label{pbeqmulti_a1b0_288_G500a}\end{figure}

As alluded to above, there are several other aspects of NEMPI that can
be compared with what has been found earlier.
We now compare our results with Fig.~6 of \cite{BGJKR14}, where the
vertical dependence of the maximum field in the structures, $B_{\max}(z)$,
was plotted, normalized either by $B_0$ or by $\Beq(z)$.
The corresponding result for our present simulations is shown in
\Fig{pbeqmulti_a1b0_288_G500a}.
The local maxima in $B_{\max}(z)$ are caused by the spatial wave-like
structures seen in \Figs{pslice1700_1770}{puz_tz_1700_1900}.

Unlike the earlier work for isothermal layers, where the slope of $\Beq/B_0$
was constant, it varies in the present case.
More importantly. the magnetic field drops significantly near the
surface and does not cross the $\Beq/B_0$ line.
This means that, unlike the earlier work with imposed vertical fields
\citep{BGJKR14}, the field in the vertical flux tubes never exceeds $\Beq$.

The magnetic field strengths of the flux concentrations are obviously
much weaker than what is expected for the Sun.
More surprising is perhaps the fact that they are also much weaker
than in the earlier isothermal models.
For $200\G$, the ratio $B_{\max}/B_0$ reaches 1.1, while
for $500\G$, it reaches 0.94. In the isothermal
case, this value could easily reach 50. We can also observe that, when
we increase the external field, $B_{\max}$ becomes smaller.

\begin{figure}[t!]\begin{center}
\includegraphics[width=\columnwidth]{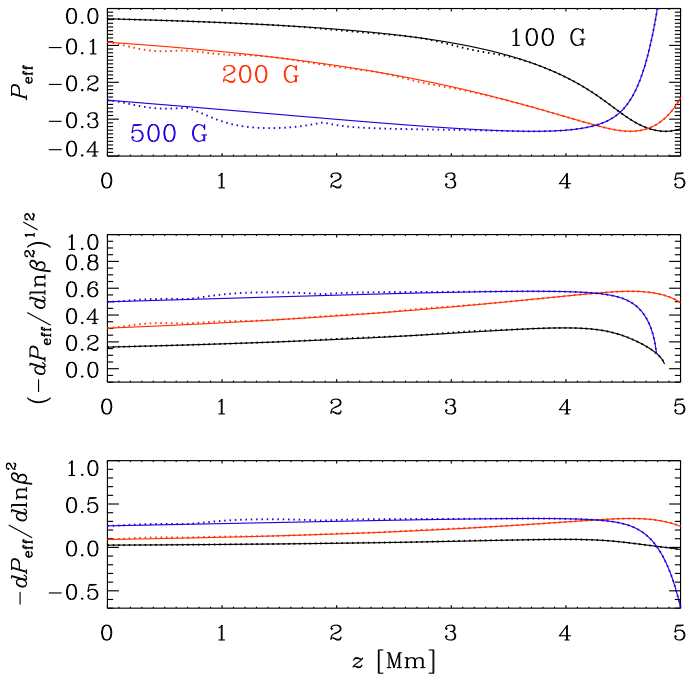}
\end{center}\caption[]{
Effective magnetic pressure 
and its derivative with respect to the magnetic field strength
for $B_0=100\G$ (black), $200\G$ (red), and $500\G$ (blue),
corresponding to $B_0/\Beqz=0.01$, $0.03$, and $0.07$.
The solid lines are based on using just the imposed magnetic field,
$\beta_0$, while the dotted lines are based on the actual field.
}\label{dpeffdlnbeta2}\end{figure}

\subsection{Effective magnetic pressure}

In \Fig{dpeffdlnbeta2} we plot the normalized effective magnetic pressure,
\EQ
\Peff(\beta)=\half[1-\qp(\beta)]\beta^2,
\EN
versus $z$.
We compare this with Fig.~9 of \cite{LBKR14}, which was a polytropic
run with $\gamma=5/3$.
In the present work, the values of $\Peff$ are about ten times larger
than in the earlier polytropic models.
This is probably related to the rather large values of $\qpz$ and
$\betastar$.
However, the shapes of the curves are similar in those two models.
Our values of the relative strength of the imposed field are similar:
for $B_0=200\G$ we have $B/\Beq=0.05$, which is comparable to the value of
\cite{LBKR14}.
Our value of $100\G$ corresponds to their ratio 0.01,
while $500\G$ corresponds to 0.07.
The results change slightly when replacing $\beta_0\equiv B_0/\Beq$ by the
value for the actual magnetic field $\beta=|\BB|/\Beq$.
Furthermore, the changes in the effective magnetic pressure caused
by the induced magnetic field are rather strong; see the dashed lines
in \Fig{dpeffdlnbeta2}.
We also see regular variations in the vertical direction, which are
associated with corresponding (time-dependent) extrema in the actual
magnetic field.

\begin{figure}[t!]\begin{center}
\includegraphics[width=\columnwidth]{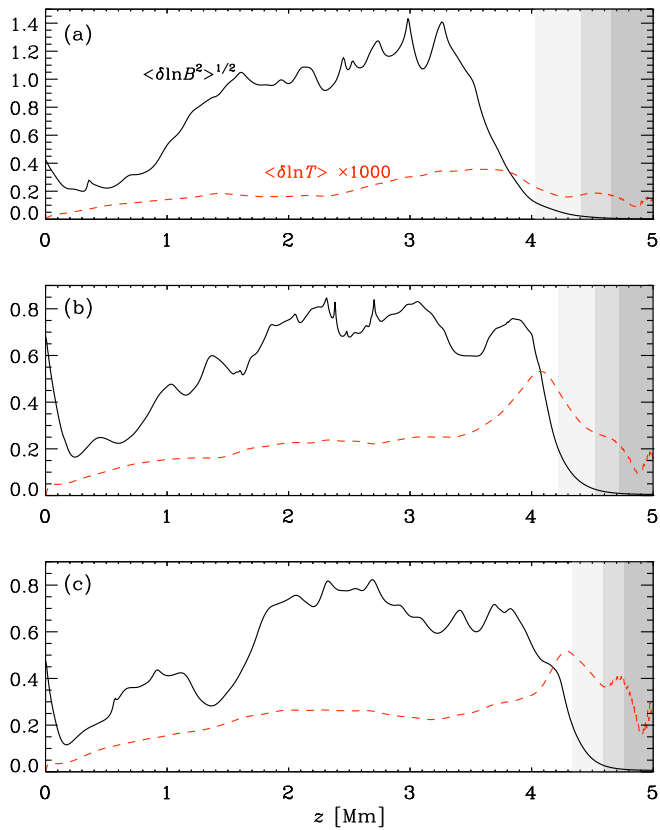}
\end{center}\caption[]{
Comparison of the vertical profiles of $\bra{\delta\ln\BB^2}^{1/2}$
and $\bra{\delta\ln T}$ (scaled by a factor 1000) for runs with
different values of $\kappa_0=2\times10^7\Mm^{-1}\cm^3\g^{-1}$ (a),
$5\times10^7$ (b), and $10^8\cm^3\s^{-1}\Mm^{-1}$ (c).
The $\tau=10$, 1, and 0.1 surfaces are indicated in gray
(from left to right).
}\label{dlnf2m_comp}\end{figure}

\subsection{Dependence on $\kappa_0$}

Increasing $\kappa_0$ means decreasing the radiative diffusive in the
deeper parts, which tends to let NEMPI appear sooner and grow faster.
It also reduces the temperature near the top of the surface and
therefore also the density scale height $H_{\rho 0}$.

To see whether radiation has a noticeable effect on NEMPI, we
compare in \Fig{dlnf2m_comp} vertical profiles of the relative
magnetic and temperature fluctuations $\bra{\delta\ln\BB^2}^{1/2}$
and $\bra{\delta\ln T}$ for runs with different values of
$\kappa_0=2\times10^7$, $5\times10^7$, and $10^8\Mm^{-1}\cm^3\g^{-1}$.
The effect is surprisingly small.
The magnetic fluctuations are of the order of unity
(and somewhat larger for $\kappa_0=2\times10^7\Mm^{-1}\cm^3\g^{-1}$),
while the relative temperature fluctuations are at most $5\times10^{-4}$.

The vigor of the temporal variation of the field increases considerably
as we increase $\kappa_0$, even though the relative strength of the
variations and the effect on the temperature remain comparable.

\begin{figure*}[t!]\begin{center}
\includegraphics[width=\textwidth]{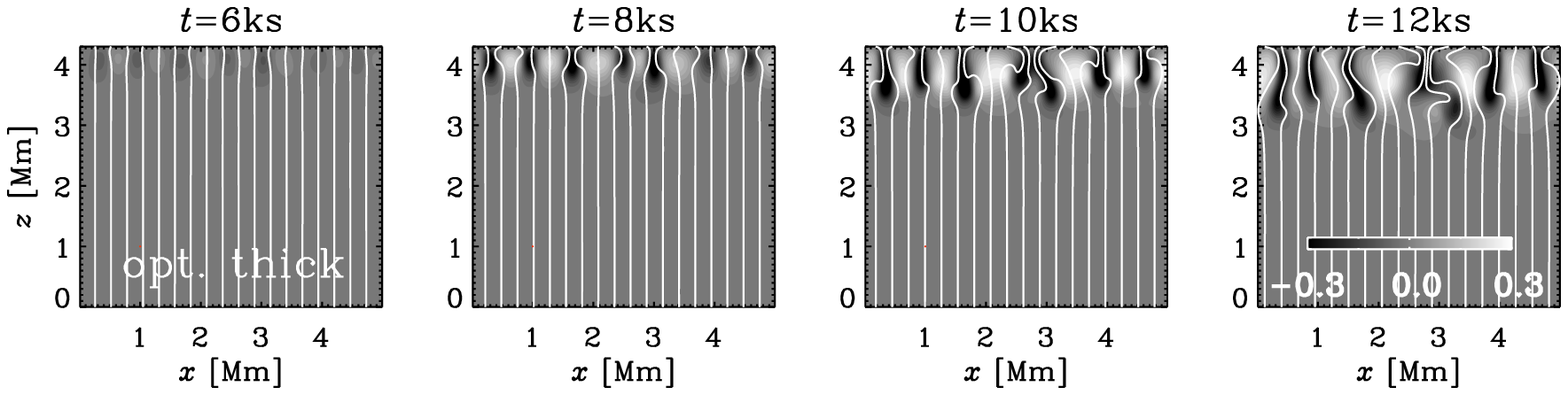}
\end{center}\caption[]{
Gray scale representation of vertical velocity
together with magnetic field lines in white
for the optically thick model at four times around
saturation of NEMPI with otherwise the same
parameters as Run~B'' with $B_0=200\G$.
}\label{pslice_thick}\end{figure*}

\subsection{Dependence on $\qpz$ and $\betastar$}

When the value of $\qpz$ is below 250, keeping $\betap=0.05$ fixed,
so $\betastar=079$,
NEMPI is found to be no longer excited and thus no magnetic structures
are created.
This remains true even when we increase $\kappa_0$ to
$5\times10^7\Mm^{-1}\cm^3\g^{-1}$,
which is generally more favorable to the onset of NEMPI.
This may indicate that there is a threshold for
$\betastar$ for the excitation of NEMPI in the presence of radiation,
which would be somewhere between 0.7 and 0.8.

\section{Comparison with simpler models}

To trace the origin of the difference to earlier results, we compare with
models without radiative transfer.
The next closest to those fully radiative models is that described in
\Sec{OpticallyThick}, in which the dynamics is optically thick, but
a radiative boundary condition \eq{radbc} is adopted at the top.
The height where this condition is applied is $z=d$,
which corresponds to the position where $\tau=1$ in the fully
radiative model; see \Fig{pstrat_a1b0_288a}.
This is at $z=d=4.3\Mm$, where the mean-free path is $\ell=0.14\Mm$,
so structures that are smaller than that experience reduced radiative
heat exchange with the surroundings in the fully radiative model, but
not in the optically thick treatment.

Another type of simplified model is one where $\means=\const$ in space
and time.
This is a strictly isentropic case, where \Eq{DsDt} is ignored.
Other than that, it has the same height and density stratification as
both the optically thick model and the fully radiative one.

\subsection{Optically thick case}

To shed some light on the occurrence of small horizontal length scales of
NEMPI in our radiative transfer models, we now compare with the optically
thick approximation discussed in \Sec{OpticallyThick}.
The result is shown in \Fig{pslice_thick} for a model that is
comparable to Run~B with $\kappa_0=10^7$.
In that case, \Eqss{ddef}{calK} yield $d=4.3\Mm$, $T_1=4998\K$, and
${\cal K}=2.7\times10^{-4}$.
It turns out that structures now develop at $z\approx4\Mm$, which is
close to the top of the domain; see \Fig{pslice_thick}.
With radiative transfer, by comparison, structures typically develop
deeper down at $z\approx3\Mm$.
However, the structures still have very small length scales comparable
to those in the models with radiative transfer.
By comparing with \Fig{pslice1700_1770} is is evident that in the models
with optically thin radiative transfer, the formation of structures at
$z\approx4\Mm$ appears to be  suppressed.
The mean free path is only about $\ell=0.14\Mm$
for our structures with $k/k_1=4$; see the red dashed line in
\Fig{pHrho_a1b0_288a}{a}.
This is rather small and can therefore not be an explanation for the
suppression of structures in the models with optically thin radiative
transfer.
There is, however, another difference between the models with optically
thin radiative transfer and the optically thick approximation that does
not have to do with NEMPI.
All models with optically thick radiative transfer have a stably stratified
layer at the top, where the entropy increases with height.
Therefore, a downdraft pulls with it high entropy material, contrary to
the case with a radiative boundary condition at $z=d$, where downdrafts
always have low entropy.
This difference was already noted by \cite{BB14}.
It explains why NEMPI does not develop near the $\tau=1$ surface at
$z=4.3\Mm$ in the optically thin radiative transfer model.
However, it does not explain the small size of NEMPI structures.
We should also point out here that, in the optically thick model, NEMPI
is no longer oscillatory.

\begin{figure*}[t!]\begin{center}
\includegraphics[width=\textwidth]{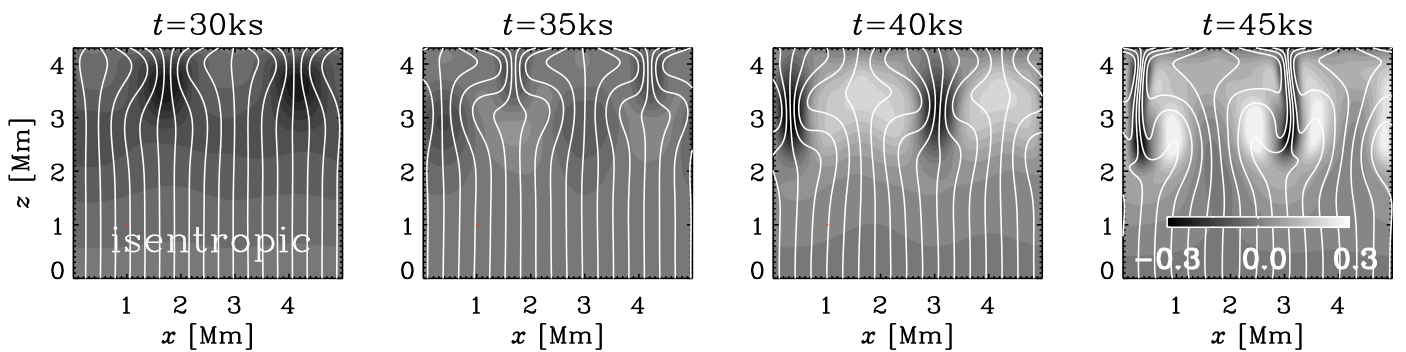}
\end{center}\caption[]{
Similar to \Fig{pslice_thick}, but for the isentropic model.
Note that also the color bar is unchanged.
}\label{pslice_adiab}\end{figure*}

\subsection{Isentropic case}

In \Fig{pslice_adiab}, we show the same model as in \Fig{pslice_thick},
but now with fixed mean specific entropy, so $\means=\const$, i.e.,
\Eq{DsDt} is not solved.
This means that the negative buoyancy is just the result of the negative
effective magnetic pressure, without any influence from changes in
specific entropy and temperature.
By contrast, when temperature and entropy are allowed to change, this
can either enhance or diminish the effect of NEMPI.
The answer discussed below is not completely straightforward.

In a stratified layer, a downdraft, even if it is initiated by NEMPI
(instead of thermal buoyancy, for example),
will always be compressed, so its density increases.
This leads to adiabatic heating, and the corresponding radiation causes a
loss of entropy, so those structures become even more negatively buoyant.
This happens most efficiently at the scale of the photon mean free path
or at the radiative diffusion scale.
Both scale are rather small and this might explain the observed tendency
for developing small structures in our model.
At the same time, however, those small length scales also make NEMPI
less efficient.
In this sense, radiation both promotes NEMPI by enhancing buoyancy
effects (both negative and positive ones), but it also counteracts NEMPI,
because it operates on progressively smaller length scales.

\section{Conclusions}

We have presented here the first calculations of NEMPI with radiation.
Within the limitations of our simplified model, NEMPI would not have
been excited had we chosen the previously determined control parameters
for the negative effective magnetic pressure effect, i.e., $\betastar$
and $\betap$.
By using a nearly three times larger value of $\betastar$,
we were able to study the reason behind this.
It turned out that in our model with radiation, the horizontal wavelength
of the instability is dramatically decreased.
As a consequence, turbulent and radiative diffusion have much stronger effects,
suppressing therefore the instability.
Nevertheless, even with a strongly enhanced value of $\betastar$, the
resulting magnetic structures are still far too weak to form sunspots.

We found for the first time that NEMPI can display oscillatory behavior
during the linear phase of the instability.
These oscillations are associated with travelling waves moving upward with a
speed of $0.2\km\s^{-1}$.
The oscillations have a period of about $4$--$9\ks$ in the volume-integrated
velocity, but since the period of the actual (signed) magnetic field is twice
as long, so the recurrence time of pronounced downward flows is $8$--$18\ks$.

We do not yet know enough about the nature of the oscillations
and whether they could also exist in reality.
To address this question further, we have to focus on the limitations
associated with the small horizontal length scales of NEMPI in the
presence of radiation.
Given that the oscillations occur only in the presence of a stably
stratified layer above, it is possible that they are related to buoyancy
oscillations in a thin upper radiative layer, where the stratification is sufficiently
stable, while still being coupled to NEMPI in the deeper layers through
suction along magnetic field lines.

The treatment of turbulent magnetic diffusion as a multiplicative factor
in front of a Laplacian diffusion operator becomes invalid on small
length scales, so the actual diffusion will be smaller; see \cite{BRS08}.
It is also possible that the opacity is still not large enough, and therefore
the radiative diffusivity is too large.
This is another unrealistic limitation of our present model.
On the other hand, in the deeper layers, the radiative diffusivity is
already now smaller than the turbulent magnetic diffusivity.
One would therefore not have expected this to be the limiting factor.
Most important is perhaps the limitation associated with the neglect of
turbulent convection in the deeper parts.
Convection would imply the presence of a strongly negative entropy
gradient just below the surface.
Therefore, the stabilizing effect from the top layers encountered in
the present model would be absent.
However, NEMPI would still lead to small length scales, except that now
turbulent convection leads to an effective thermal diffusivity that is
much larger than the radiative one.
Moreover, the transition between a radiative surface above and strong
turbulence with small-scale convection beneath the surface would be
very abrupt.
Given that NEMPI is most effective for large scale separation
(small-scale turbulence) and the stratification is strongest near the
surface, it might still be a viable alternative for the formation of
sunspots.
Extending our model by including convection in parameterized form would
therefore be a first task to be addressed in a follow-up investigation.

Ultimately, the aim is to model the formation of sunspots, where convective
heat transport is either suppressed by the magnetic field \citep{Bie41}
or the cooling enhanced \citep{Par74}.
The former effect may lead to its own instability, which
was modelled by \cite{KM00} using a mean-field approach.
This instability could be strengthened further by the effects of ionization
and would therefore be another urgent target for subsequent investigations.

\begin{acknowledgements}
We thank the referee for useful comments and
Sacha Brun for support and encouragement.
Support through the NSF Astrophysics and Astronomy Grant Program (grant
1615100) and the Research Council of Norway (FRINATEK grant 231444) are
gratefully acknowledged.
We acknowledge the allocation of computing resources provided by
the Swedish National Allocations Committee at the Center for Parallel
Computers at the Royal Institute of Technology in Stockholm.
This work utilized the Janus supercomputer, which is supported by the
National Science Foundation (award number CNS-0821794), the University
of Colorado Boulder, the University of Colorado Denver, and the National
Center for Atmospheric Research.
The Janus supercomputer is operated by the University of Colorado Boulder.
\end{acknowledgements}


\vfill\bigskip\noindent\tiny\begin{verbatim}
$Header: /var/cvs/brandenb/tex/barbara/meanNEMPI/paper.tex,v 1.86 2017/09/20 09:05:47 brandenb Exp $
\end{verbatim}

\end{document}